# Reliability of spin-to-charge conversion measurements in graphene-based lateral spin valves

C. K. Safeer[1,2], Franz Herling[1], Won Young Choi[1], Nerea Ontoso[1], Josep Ingla-Aynés[1], Luis E. Hueso[1,3], Fèlix Casanova[1,3,*].

[1] CIC nanoGUNE BRTA, 20018 Donostia-San Sebastian, Basque Country, Spain.
[2] Department of Physics, Clarendon Laboratory, University of Oxford, Oxford, UK.
[3] IKERBASQUE, Basque Foundation for Science, 48013 Bilbao, Basque Country, Spain.

*E-mail: f.casanova@nanogune.eu

## Abstract

Understanding spin physics in graphene is crucial for developing future two-dimensional spintronic devices. Recent studies show that efficient spin-to-charge conversions via either the inverse spin Hall effect or the inverse Rashba-Edelstein effect can be achieved in graphene by proximity with an adjacent spin-orbit coupling material. Lateral spin valve devices, made up of a graphene Hall bar and ferromagnets, are best suited for such studies. Here, we report that signals mimicking the inverse Rashba-Edelstein effect can be measured in pristine graphene possessing negligible spin-orbit coupling, confirming that these signals are unrelated to spin-to-charge conversion. We identify either the anomalous Hall effect in the ferromagnet or the ordinary Hall effect in graphene induced by stray fields as the possible sources of this artefact. By quantitatively comparing these options with finite-element-method simulations, we conclude the latter better explains our results. Our study deepens the understanding of spin-to-charge conversion measurement schemes in graphene, which should be taken into account when designing future experiments.

## Introduction

Graphene is an outstanding material for long-distance coherent spin transport [1–4]due to its weak intrinsic spin-orbit coupling (SOC) and negligible hyperfine interaction. For the same reason, it is not the preferred material for active spintronics device applications which require strong SOC. However, theoretical studies suggested that SOC can be induced in graphene via either proximity by combining it with materials possessing large SOC [5–12] or adatom decoration [13–16]. As a consequence, spin-orbit effects such as the direct or inverse spin Hall effect (SHE and ISHE) [17] and the Rashba-Edelstein effect (REE and IREE) [18] have been predicted theoretically [6,7,9,11,13–16]. Experimentally, an efficient spin-to-charge conversion (SCC) due to ISHE was first unequivocally observed in graphene/$MoS_2$ van der Waals heterostructures [19]. Later, SCC was reported in different systems of graphene combined with other transition metal dichalcogenides (TMDs) [20–25], layered topological insulators [26] or insulating bismuth oxide (Bi$_2$O$_3$) [27].

In all the SCC studies in graphene published so far [16–24], a graphene-based lateral spin valve (LSV) device has been used for the measurements. Graphene-based LSVs have also been used to measure SCC in other SOC materials [25–30]. The basic measurement layout is shown in figures 1c-1h: An electrical current injected across a ferromagnet/graphene interface creates a spin accumulation which then diffuses as a spin current towards the proximitized graphene region. There, SCC leads to the creation of a transverse charge current, thus a corresponding non-local output voltage ($V_{NL}$) is measured across the graphene/SOC material Hall bar. The two SCC mechanisms in proximitized graphene [19], ISHE and IREE result in the conversion of out-of-plane (figure 1a) and in-plane spins (figure 1b), respectively. They can be differentiated by performing the SCC measurements in the presence of an in-plane ($B_x$) or out-of-plane ($B_z$) magnetic field as shown in figures 1c-1h. Initially, the magnetization of the ferromagnet and therefore the polarization of the injected spins are aligned along its easy axis ($y$). Then, a magnetic field $B_x$ ($B_z$) is applied along the hard axis and causes a change in the polarization of the spins arriving at the Hall bar region in two ways. Firstly, at low magnetic fields, the spins precess in the $y-z$ ($y-x$) plane during their diffusion[35,36] in the graphene channel as shown in figure 1c-d. Hence, the polarization of the spins arriving at the Hall bar region rotates, giving rise to oscillations in $V_{NL}$. Because the converted spins are perpendicular to the injected ones, $V_{NL}$ becomes an antisymmetric Hanle spin precession curve $vs. B_x$ ($B_z$) as shown in figure 1e. This Hanle precession curve reverses its sign if the initial easy axis magnetization of the ferromagnet is reversed. Secondly, as the magnetic field gets stronger, the magnetization of the ferromagnetic electrode gets pulled and subsequently saturated towards the hard axis, correspondingly changing the polarization of the injected spins. $B_x$ and $B_z$ tilt the injected spins along the $x$ (figure 1f) and $z$ axis (figure 1g), respectively, resulting in an S-shaped $V_{NL}$ $vs. B$ curve as schematically shown in figure 1h. This curve is independent of the initial orientation of the ferromagnet along the easy axis (as seen for the superposed red and blue curves). In summary, if an in-plane SCC is observed obtaining an S-shaped $V_{NL}$ $vs. B_x$ curve due to contact pulling (figure 1f) or/and an antisymmetric Hanle spin precession (figure 1d) $V_{NL}$ $vs. B_z$ curve, the SCC mechanism is considered to be IREE in proximitized graphene [17–20]. It is also worth noting that SCC can also occur via ISHE in the SOC material on top of graphene, which also results in in-plane SCC [19,20,28,29,31–34], making it hard to distinguish from IREE in graphene. However, if an out-of-plane SCC is measured by obtaining an antisymmetric Hanle $V_{NL}$ $vs. B_x$ curve (figure 1c) or/and an S-shaped $V_{NL}$ $vs. B_z$ curve (figure 1g), the SCC is reported to be due to ISHE in proximitized graphene [19,21,22,24,27].

In this article, we report that an S-shaped $V_{NL}$ $vs. B$ curve can be obtained both in pristine graphene as well as in graphene combined with a high-SOC material (Bi$_2$O$_3$). Therefore, this observation is unrelated to SCC in graphene arising from any SOC phenomenon, and most likely due to measurement artefacts related to the variation of the magnetization of the ferromagnet. We conclude that the only unambiguous method to measure SCC using graphene-based lateral spin valves is the spin precession measurement.

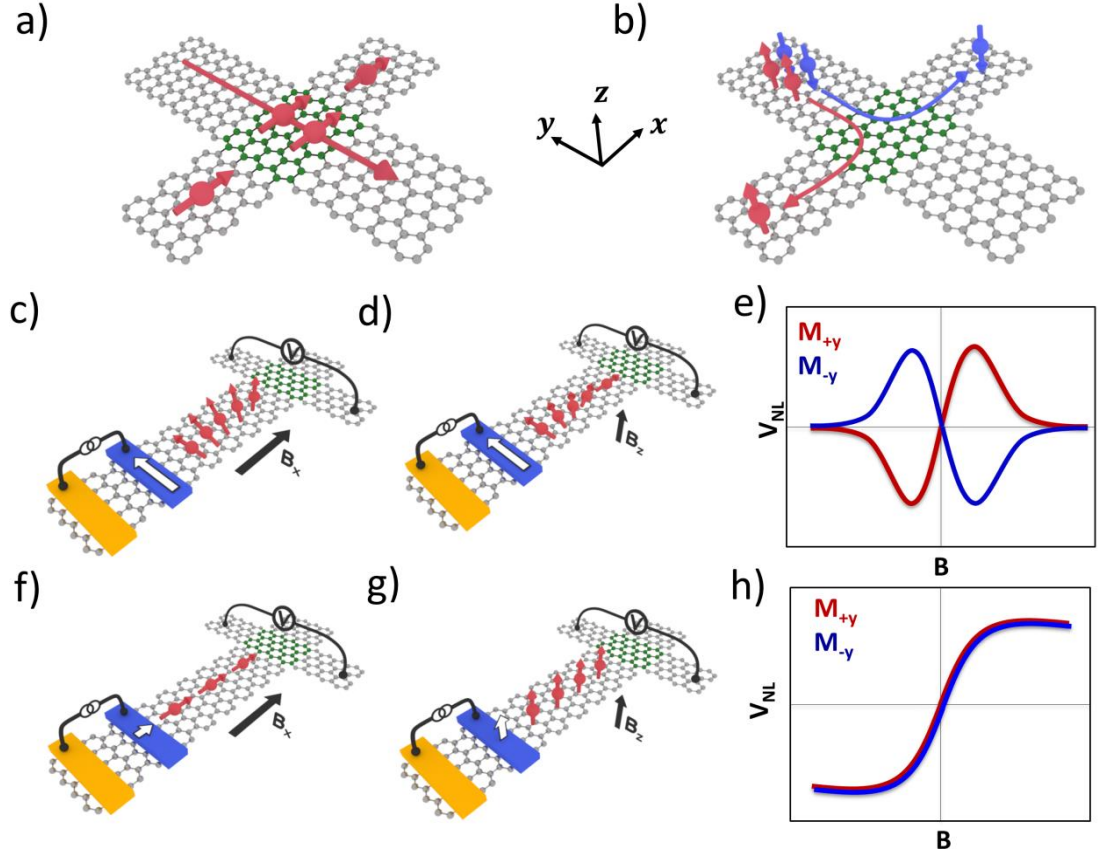

**Figure 1.** **a)** Schematic diagram of the REE in proximitized graphene. The green region represents graphene proximitized with an adjacent SOC material (not shown). An electrical current applied along $y$ creates a spin accumulation with polarization along $x$ at the proximitized region which then diffuses in the graphene channel. **b)** Schematic diagram of the SHE in proximitized graphene. An electrical current applied along $y$ creates a spin current diffusing along $x$ with spin polarization along $z$. **c)** SCC measurement scheme. The spins are injected by applying a current between the ferromagnet and graphene and SCC is measured by probing $V_{NL}$ across the graphene Hall bar. $V_{NL}$ is due to spin precession in the $y - z$ plane if the magnetic field is applied along the $x - $ axis ($B_x$) and **d)** in the $x - y$ plane if the magnetic field is applied along the $z - $ axis ($B_z$). **e)** The expected antisymmetric Hanle spin precession curves for the cases explained in panels c and d. The red and blue curve corresponds to the initial magnetization of the ferromagnetic electrode set along $+y$ axis and $-y$ axis, respectively. **f)** SCC due to the variation of the spin polarization via pulling the magnetization of the ferromagnetic electrodes with the magnetic field applied along $x - $ axis ($B_x$) and **g)** $z - $ axis ($B_z$). **h)** The expected $V_{NL}$ vs. $B$ curve for the cases explained in panels e and f. The red and blue curve corresponds to the initial magnetization of the ferromagnetic electrode set along $+y$ axis and $-y$ axis, respectively.

## Device fabrication

Figure 2a shows the scanning electron microscope image of the used device. At first, a bilayer graphene (BLG) flake was exfoliated onto a highly doped Si substrate covered with SiO$_2$. The number of layers was identified according to the optical contrast and the Raman spectrum [37] (See Supplementary Information S1). Then, using electron-beam lithography and reactive ion etching, the graphene flake was shaped into a Hall bar with three crosses. Subsequently, using electron-beam lithography followed by thermal and electron-beam evaporation, Cr (5 nm)/Au (40 nm) contacts were connected to the graphene Hall bar. Using similar steps, Bi$_2$O$_3$ (5 nm) was deposited in the middle of two Hall bars and several TiO$_x$/Co (35 nm) ferromagnetic electrodes were placed on top of the graphene channel to create LSVs. TiO$_x$ (~0.3 nm) barriers were used to ensure efficient spin injection into graphene [38]. The widths of the ferromagnetic electrodes are alternating (~450 nm and ~300 nm) so that they have different coercive fields due to different shape anisotropy, enabling parallel and antiparallel configurations by applying a magnetic field along their easy axis.

In the manuscript, we focus on the measurements performed at T= 50 K. Room temperature measurements are shown in Supplementary Information S4. To verify the reproducibility of our results, similar measurements were performed in the left side (figure 2a) of the same device (Device 1, Supplementary Information S5) and on additional devices (Device 2 and 3, Supplementary Information S6 and S7 respectively).

## Results

**Spin transport measurements**

In the first step, the spin transport parameters of the pristine graphene were extracted. For this, the typical Hanle spin precession measurement using a conventional four-terminal non-local geometry [1] was performed. An electrical current ($I_c$) of 10 $\mu$A was applied between the Co electrode 4 and Au electrode E and the non-local voltage ($V_{NL}$) was measured between Co electrode 3 and Au electrode A, from which the corresponding non-local resistance ($R_{NL} = \frac{V_{NL}}{I_c}$) was calculated. Initially, the magnetizations of the ferromagnets were set to the parallel and antiparallel configurations by applying $B_y$. Then, for each initial state, $B_x$ was swept from 0 to 0.6 T and 0 to -0.6 T and the corresponding variations in $R_{NL}$ ($R_{NL}^P$ and $R_{NL}^{AP}$) were obtained as shown in figure 2b. Subsequently, the contributions to $R_{NL}$ due to pure spin precession (Supplementary Information S10, figure S10a) and the rotation angle of the Co magnetization (figure 2c) were disentangled as explained in Ref. [19]. The pure spin precession curve was then fitted using the solutions of the Bloch equation [19,39], obtaining the spin lifetime of graphene ($\tau_s^{gr}$) ~ 92±10 ps, the spin diffusion constant of graphene ($D_s^{gr}$) ~ (10.1±1) × 10$^{-3}$ m$^2$/s, and the spin polarization at the Co/graphene interface ($P$) ~ 4.4±0.1 %. The details of the fitting are explained in Supplementary Information S10. A similar measurement and analysis were performed using the LSV made up of Co electrodes 2 and 3 (figure 2c) where the graphene channel consists of the Hall bar region with Bi$_2$O$_3$ deposited in the middle, which is also explained in Supplementary Information S10.

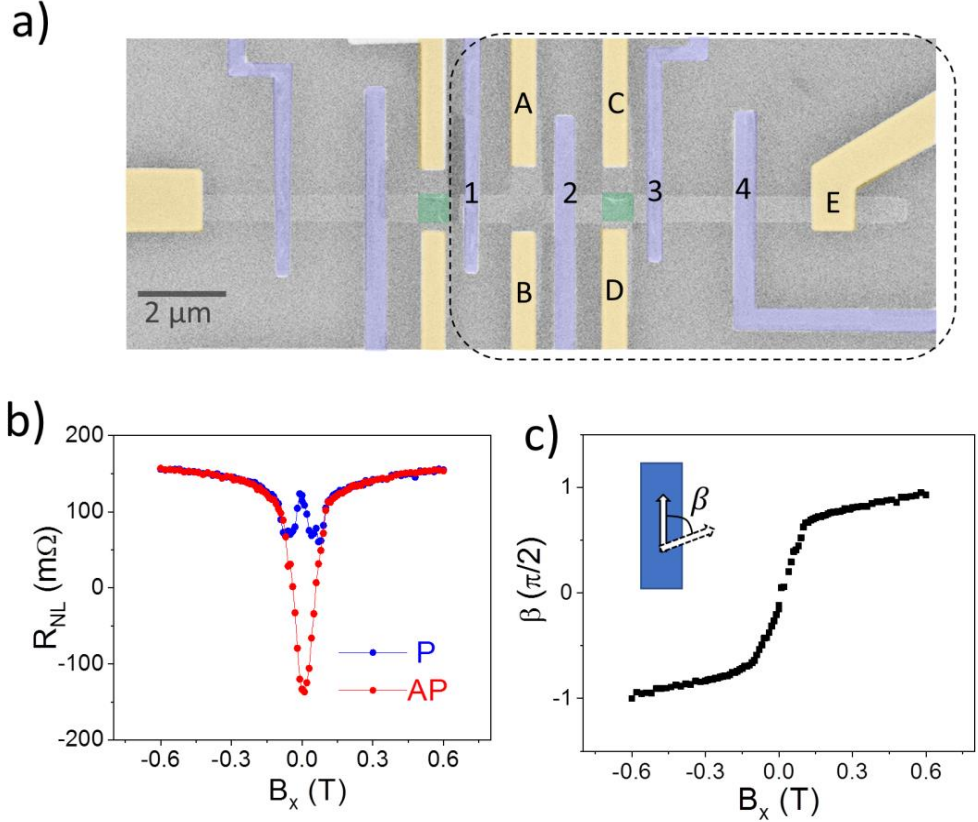

**Figure 2. a)** False coloured scanning electron microscope image of the graphene LSV device measured. The Cr/Au electrodes, TiO$_x$/Co electrodes, and Bi$_2$O$_3$ regions are highlighted in yellow, purple, and green, respectively. The measurements were performed in the device region highlighted by the dashed box. **b)** Symmetric Hanle curves measured at 50 K and $V_g = 10$ V for initial parallel (blue curve) and antiparallel (red curve) states of the two ferromagnets across the pristine graphene region, using electrical configuration V$_{3A}$I$_{4E}$. **c)** $\beta$, the angle between the magnetization and its easy axis (**y**), as a function of $B_x$ extracted from the measurement in panel b. The details of the analysis are explained in the Supplementary Information S10.

**Non-local measurements using spin-to-charge conversion geometry**

After confirming an efficient spin transport in our device, we performed SCC measurements in the graphene/Bi$_2$O$_3$ region. For this, $I_c$ was applied between Co electrode 2 and Au electrode A and $V_{NL}$ was measured using Au electrodes C and D across the graphene/Bi$_2$O$_3$ Hall bar (figure 2a). As plotted in figure 3a, an S-shaped $R_{NL}$ vs. $B_x$ curve was obtained. This measurement may indicate SCC of in-plane spins (see figures 1f and 1h), hence the possibility of IREE in graphene/Bi$_2$O$_3$. Since our previous study shows that ISHE can occur in graphene/Bi$_2$O$_3$ due to proximity-induced SOC [27], obtaining IREE in the same system is a plausible scenario. To further investigate this option, we performed a control experiment across the pristine graphene region by applying $I_c$ between Co electrode 2 and Au electrode E and $V_{NL}$ was measured using Au electrodes A and B. Interestingly, as shown in figure 3b, we again observe the S-shaped $R_{NL}$ vs. $B_x$ curve, saturating at the same field values, but with an amplitude even larger than that of the graphene/Bi$_2$O$_3$ region. Since pristine graphene possesses negligible SOC, we conclude that the measured signal must have a different origin, unrelated to SCC.

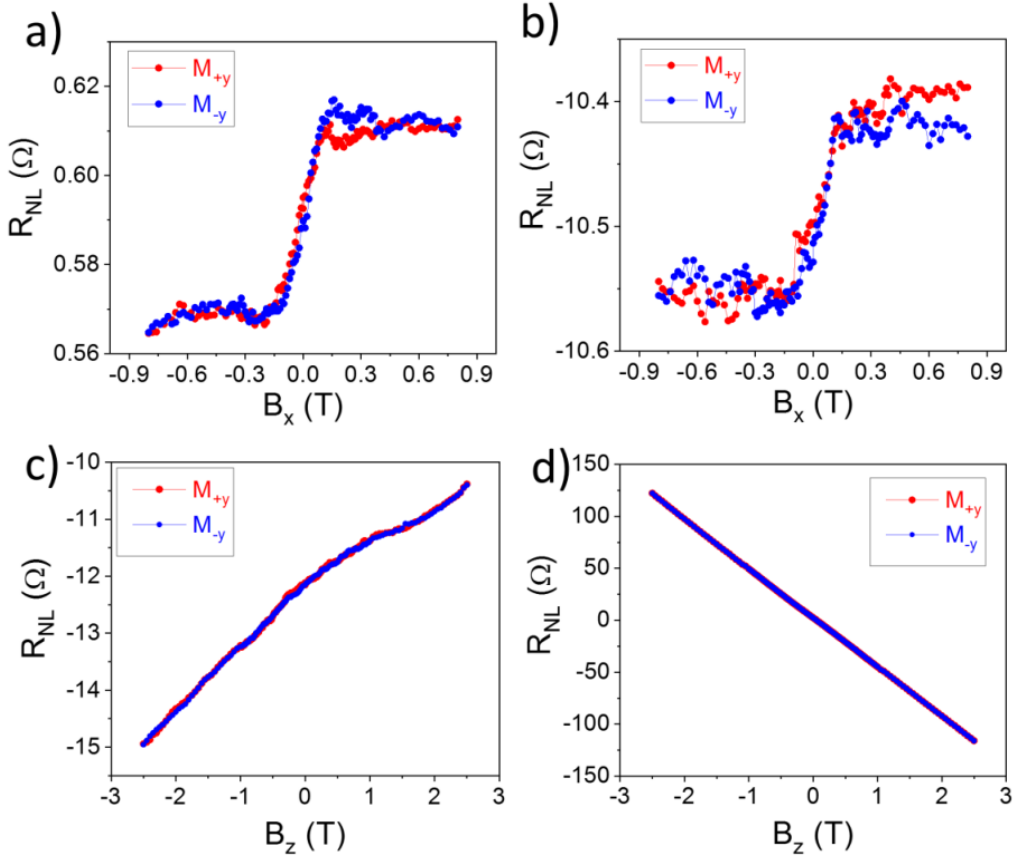

**Figure 3.** $R_{NL}$ $vs.$ $B_x$ measurements across **a)** graphene/Bi$_2$O$_3$ using electrical configuration $V_{CD}I_{2A}$ and **b)** pristine graphene using electrical configuration $V_{AB}I_{2E}$. $R_{NL}$ $vs.$ $B_z$ measurement across **c)** graphene/Bi$_2$O$_3$ and **d)** pristine graphene. The red and blue curves correspond in all panels to the initial magnetization of the Co injector saturated along the $+y$ and $-y$ easy axis directions, respectively. All measurements are taken at 50 K.

Even though we can conclude that the signal observed across pristine graphene is not related to SCC, the $R_{NL}$ $vs.$ $B_x$ curve obtained across the graphene/Bi$_2$O$_3$ region can still be due to either artefacts alone or a mixture including SCC by IREE. An easy way to test the possibility of in-plane SCC by IREE is to induce spin precession by applying $B_z$ (figure 1d) in which an antisymmetric Hanle curve (figure 1e) is expected. Since the spin precession occurs in the graphene channel and is separated from the variation of the Co magnetization, any artefact arising from the ferromagnet can be avoided and only pure spin-related phenomena will be detected. The amplitudes of the signals in figures 3a and 3b are in the order of mΩs and, if they arise from SCC, a similar or smaller amplitude is expected for the spin precession signal. However, as shown in figures 3c and 3d, a large and linear $R_{NL}$ $vs.$ $B_z$ signal (in the order of Ωs) was observed, which most likely hinders detection of any spin precession signal. It is well known that, even in non-local geometries, a small spurious current can diffuse from the injector acquiring a $y$-component in the graphene channel [40]. The linear signal observed here is most likely the spurious current-induced ordinary Hall effect (OHE) in graphene in the presence of $B_z$. Due to the large OHE in graphene [41], a small spurious current can create a local voltage large enough to dominate over the non-local SCC voltage. Since such small charge current spreading is unavoidable, we conclude that it is not possible to detect spin precession using $B_z$ in our samples. However, the spin precession using $B_z$ would be measurable [20,22,33,34] by minimizing the spurious current effects in different ways: Firstly, optimizing the graphene/oxide/ferromagnet tunnel barrier to increase spin injection while lowering the spurious current injection by improving homogeneity, which is generally achieved at larger interface resistance (~kΩs) [4,42]. Secondly, fabricating a LSV device with a narrower graphene channel width and longer distance between the Co electrode and the Hall bar. Since this also reduces the spin current reaching the detector, a compromise for the device dimensions has to be found to obtain a detectable output signal. Using graphene with better spin transport properties for example through encapsulation will make this easier. Thirdly, the design can also incorporate the recently proposed [26] LSV with an orthogonal graphene channel. Here, the in-plane spin precession can be measured by applying $B_x$ instead of $B_z$ so that the $B_z$-induced spurious effects can be minimized.

To further verify our results, the same measurements as those in figures 2 and 3 were performed in the different part of Device 1, in Device 2 and Device 3 (see Supplementary Information S5, S6 and S7). We obtained similar $R_{NL}$ $vs.$ $B_x$ and $R_{NL}$ $vs.$ $B_z$ curves, confirming the reproducibility of our results. Since Device 3 was made of monolayer graphene Hall bar, we also conclude that the spurious effect is qualitatively independent of the number of graphene layers.

## Discussion

The $R_{NL}$ $vs.$ $B_z$ curve in figure 3d does not show any saturation behaviour, while $R_{NL}$ $vs.$ $B_x$ in figure 3b saturates at ~ ±1500 Oe, the same value at which the magnetization of the Co electrode saturates (figure 2e). This indicates that the shape of the $R_{NL}$ $vs.$ $B_x$ curve is directly related to the variation of the Co magnetization. To further confirm this, we fabricated another device (Supplementary Information S8) with Co and Au electrodes placed at the right and left side of the graphene Hall bar, respectively. As shown in figure 4a, the S-shaped $R_{NL}$ $vs.$ $B_x$ curve was only obtained when the Co electrode was used as the injector, but not for the Au electrode, further confirming the dependency on the Co magnetization variation.

As reported previously [21–23,25,26,43], a possible mechanism explaining our results is the stray field-induced OHE in graphene. An out-of-plane magnetic field applied to graphene can cause OHE. As the Co magnetization tilts and saturates towards the $x$ direction, an out-of-plane stray field generated at the edges of the Co electrode through the graphene channel linearly increases and saturates. In combination with a small spurious current from the ferromagnetic injector this could lead to OHE in graphene resulting in a linearly varying and saturating $R_{NL}$ $vs.$ $B_x$ curve as we observed in figures 3a and 3b.

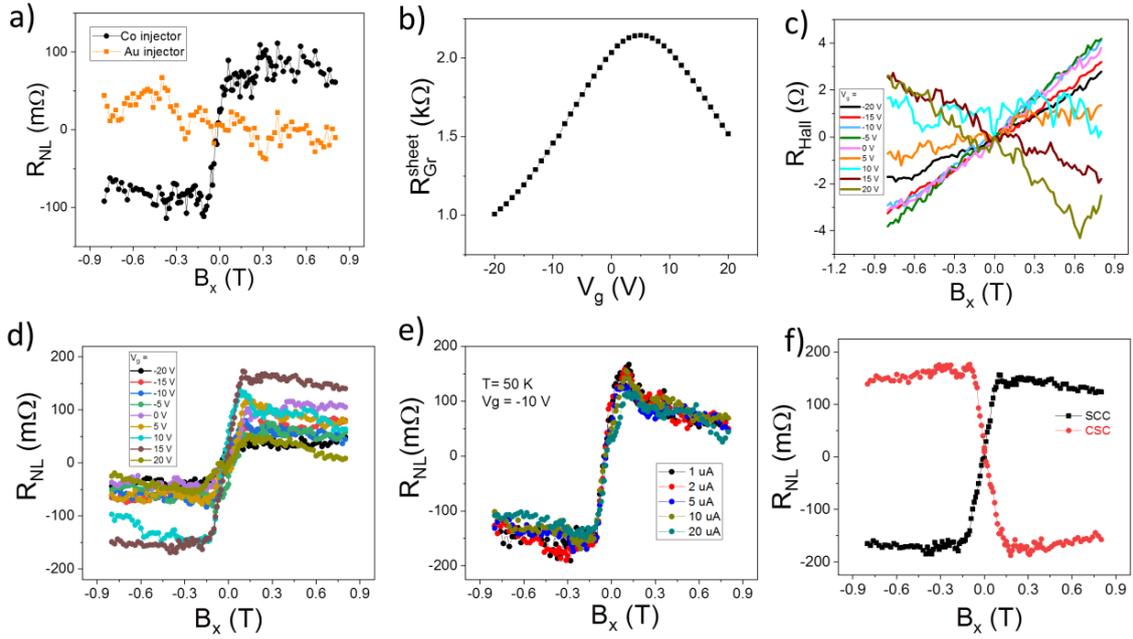

**Figure 4. a)** $R_{NL}$ vs. $B_x$ measurements using Co and Au injectors obtained using the device explained in Supplementary Information figure S8a. The S-shaped behaviour was obtained only when the ferromagnetic injector was used. **b)** Sheet resistance of graphene ($R_{Gr}^{sheet}$) as a function of $V_g$ measured using 4-probe electrical configuration V$_{34}$I$_{AE}$. **c)** The $R_{Hall}$ vs. $B_x$ measurements using electrical configuration V$_{AB}$I$_{1E}$ for different $V_g$. The $R_{NL}$ vs. $B_x$ measurements across the pristine graphene region, **d)** for different $V_g$ from -20 V to 20 V at $I_C$ = 10 μA, **e)** for $I_C$ = 1, 2, 5, 10, 20 μA at $V_g$ = -10 V, **f)** in the SCC and charge-to-spin conversion (CSC) geometries using electrical configuration V$_{AB}$I$_{1E}$ and V$_{1E}$I$_{AB}$, respectively. All measurements are taken at 50 K.

A control experiment to check the above-mentioned possibility is to perform gate voltage ($V_g$)-dependent measurements. For the hole and electron-doped regimes in graphene, which can be obtained by applying $V_g$, opposite signs for the OHE are expected and a corresponding sign change is also expected for the $R_{NL}$ vs. $B_x$ curve. To verify this possibility, we performed different measurements as a function of $V_g$ as shown in figures 4b to 4d. At first, using a 4-probe configuration (V$_{34}$I$_{AE}$), the sheet resistance of graphene was measured as a function of $V_g$ (figure 4b). The charge neutrality point was observed at $V_g$ ~ +5 V, confirming hole and electron carriers dominate transport in the channel for $V_g$ < +5 V and $V_g$ > +5 V, respectively. Then, the Hall resistance ($R_{Hall}$) as a function of $B_x$ was obtained at different $V_g$ (figure 4c) by applying the current along the graphene channel (contacts 1 and E) and measuring the voltage across the Hall bar (contacts A and B). A linearly varying $R_{Hall}$ vs. $B_x$ was observed with opposite slopes for electron and hole-doped regimes. The OHE here should be caused by a $B_z$ field component originated from a small out-of-plane tilt between the sample and $B_x$. After confirming the sign reversal of the OHE, we then performed $R_{NL}$ vs. $B_x$ measurements at different $V_g$ (figure 4d). Interestingly, unlike the previous reports, we did not observe any sign reversal. A similar $V_g$-dependent measurement was also obtained across the graphene/Bi$_2$O$_3$ Hall bar region as shown in

Supplementary Information S3. It is important to note that these measurements further confirm the absence of IREE, as such a SCC signal is also expected to change sign for the two regimes [21–23,25,26].

Two different hypotheses can explain the gate dependent studies mentioned above. Firstly, the origin of the artefact cannot be OHE at the graphene Hall bar However, if the OHE is from the graphene region near the Co electrode where the stray field is maximum and has a different doping compared to the Hall bar region, the carrier type will be unchanged with the applied gate voltages. To experimentally check whether this is the case, we fabricated a graphene Hall bar device with Co deposited on the Hall cross as explained in Supplementary Information S8. The transfer curve of the pristine graphene region showed the charge neutrality point at $V_g \sim +4V$ (figure S9b). However, the Hall measurement in the graphene/Co region did not change sign for -10 V$\leq V_g \leq$ +10 V (figure S9c) confirming the carrier type is unchanged. Secondly, due to the anomalous Hall effect (AHE) in the ferromagnetic Co electrode, a voltage along the in-plane $y$ direction can build up in Co with magnetization along $x$ [44], when a current is applied along $z$. If the adjacent graphene Hall bar probes this voltage, an S-shaped $R_{NL}$ $vs.$ $B_x$ curve could be obtained. Importantly, the sign of AHE is expected to be independent of $V_g$, explaining our gate measurements shown in figure 4d.

To quantitatively understand the contributions to the output signal from those two effects, we performed 3D Finite Element Method simulation using COMSOL Multiphysics (the details of the simulation are explained in Supplementary Information S11) as shown in figure 5. For both calculations, the experimentally measured Co (50 μΩ·cm) and graphene (162 μΩ·cm) resistivities have been used. Considering the saturation magnetization of the Co electrode ($1.4\times10^6$ A/m) [45], the $z$-component of its stray field is obtained. The stray field is maximum near the ferromagnetic edge and negligible at the Hall bar region. Then, the OHE for an electron concentration in graphene of $3.6\times10^{11}$ cm$^{-2}$ was calculated. This OHE affects the electric potential profile of the Hall bar (detector) causing a voltage difference measured across the Hall bar (figure 5a). Due to the complexity in the calculations, the minimum thickness of graphene that can be used here is limited to 8 nm. By plotting the nonlocal signal as a function of graphene thickness at a detector distance of 600 nm (the same detector distance used for the experiment with the device shown in figure 2a) and extrapolating the linear plot to the thickness of bilayer graphene (0.8 nm), a nonlocal signal of ~65 mΩ was obtained (figure 4d). This signal value is in the same order of magnitude as the experimental one. AHE also affects the electric potential as shown in figure 5b. Here, an anomalous Hall angle of 1% [46] has been used for the calculation. As shown in figure 5d, the signals for both cases decrease as the distances between the detector and the edge of the ferromagnet increase. Also, the contribution from the AHE is very small (~1 mΩ) compared to that from the OHE. Therefore, we conclude the source artefact contributing to the nonlocal measurement should be OHE, while a small contribution from the AHE may still exist.

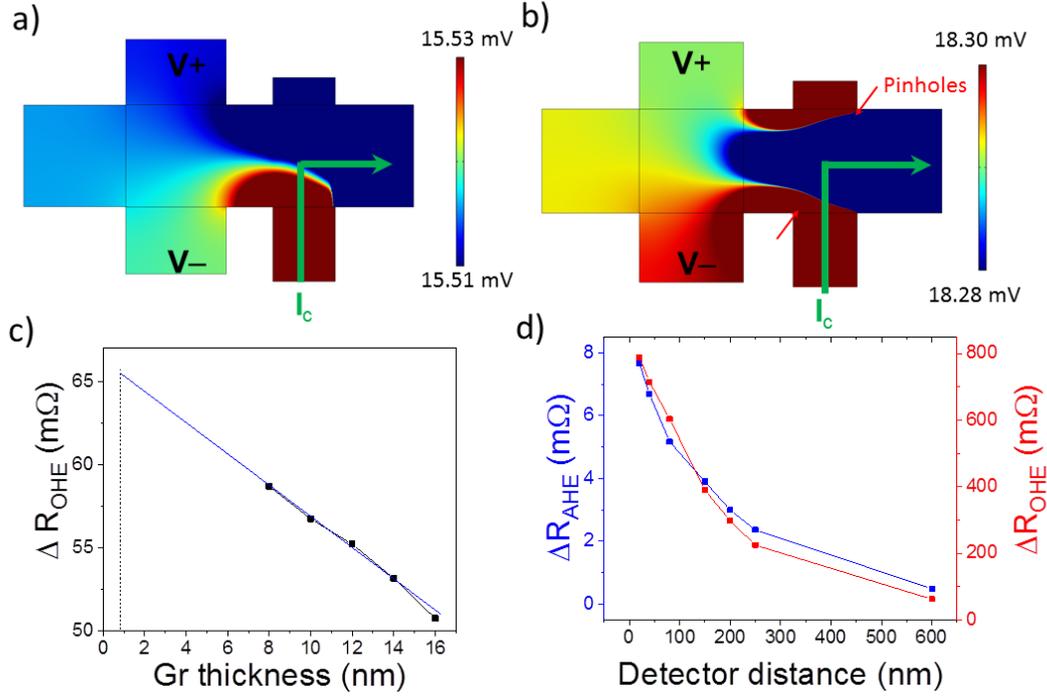

**Figure 5.** The electric potential profiles obtained from 3D Finite Element Method simulations due to **a)** stray-field-induced OHE for 8-nm-thick graphene, where a transparent ferromagnet/graphene interface has been considered, and **b)** AHE in the ferromagnetic electrode where pinholes indicated by red arrows and an insulating $TiO_x$ barrier are considered. The current of 10 $\mu A$ is applied to the lower end of the Co electrode and flows to the right end of the graphene channel (green arrow) while the voltage is measured across the Hall probes. The voltage drop following the current line is set out of range to highlight the electric potential profile at the non-local region. Further details of the simulations are explained in Supplementary Information S11. **c)** Amplitude of the OHE, measured at 600 nm from the ferromagnet edge, as a function of the graphene thickness. Since the simulation is limited to 8 nm of graphene, the linear plot is extrapolated to 0.8 nm, which is assumed to be the thickness of bilayer graphene. **d)** Amplitude of the AHE- (blue) and OHE-induced (red) non-local resistances as a function of distance between the detector (center of the Hall bar) and the edge of the ferromagnet.

Finally, we performed two experiments usually taken as evidence for SCC in these systems. Firstly, to check the dependence on the strength of the electrical current, $R_{NL}\ vs.B_x$ was measured for $I_C$ = 1, 2, 5, 10, 20 $\mu A$ (figure 4e). It shows a linear response, i.e., the amplitude of $R_{NL}$ is independent of the applied current. The linear response also leads to Onsager reciprocity as shown in figure 4f, which was measured by exchanging the current and voltage terminals. These two features were also reported to appear for SCC due to IREE [20–23,25,26,28,29,33,34]. Since the artefacts caused by the spurious current, originating from OHE or AHE, have a similar response, such measurements should not be used as confirmation of SCC in graphene LSV devices.

In summary, we performed SCC-like measurements in graphene and in graphene proximitized with a SOC material. An S-shaped $R_{NL}\ vs.B_x$ signal mimicking the previously reported SCC signal due to IREE was observed in pristine graphene where, due to the low SOC, any SCC mechanism should be negligible. Therefore, we conclude our signal is due to non-SCC-related artefacts. By performing gate-dependent measurements, we conclude that two scenarios can explain our observation: the OHE in the graphene region near the ferromagnetic electrode or the AHE in the ferromagnet. The quantification of these scenarios with 3D simulations indicates that our result can be explained by considering only OHE. Finally, we would like to point out that extreme care should be taken while using the S-shaped $R_{NL}\ vs.B$ measurements for the quantitative and qualitative interpretation of SCC using graphene-based LSVs. Only spin precession measurement should be used for unambiguous SCC claims. As in our $R_{NL}\ vs.B_z$ measurements, it might be impossible to perform spin precession if the signal due to spin-unrelated effects dominates over the SCC signal. In this case, additional efforts to minimize artefacts by carefully designing the dimensions of the graphene channel and electrodes [20,22,33,34], the ferromagnet/graphene interface to reduce the spurious current, or adapting the recently reported new geometrical design for the graphene-based LSVs [26] will be required.

## Acknowledgements

This work is supported by the Spanish MICINN under Project RTI2018-094861-B-100 and under the Maria de Maeztu Units of Excellence Programme (MDM-2016-0618), by the European Union H2020 under the Marie Slodowska Curie Actions (794982-2DSTOP and 0766025-QuESTech), and by the "Valleytronics" Intel Science Technology Center. W.Y.C. and J.I.-A. acknowledge postdoctoral fellowship support from the "Juan de la Cierva - Formación" program by the Spanish MICINN (Grants No. FJC2018-038580-I and FJC2018-038688-I, respectively). N.O. thanks the Spanish MICINN for a Ph.D. fellowship (Grant no. BES-2017-07963).

# Supplementary Information

# Reliability of spin-to-charge conversion measurements in graphene-based lateral spin valves

C. K. Safeer[1,2], Franz Herling[1], Won Young Choi[1], Nerea Ontoso[1], Josep Ingla-Aynés[1], Luis E. Hueso[1,3], Fèlix Casanova[1,3,*].

[1] CIC nanoGUNE BRTA, 20018 Donostia-San Sebastian, Basque Country, Spain.
[2] Department of Physics, Clarendon Laboratory, University of Oxford, Oxford, UK
[3] IKERBASQUE, Basque Foundation for Science, 48013 Bilbao, Basque Country, Spain

*E-mail: f.casanova@nanogune.eu

## Table of contents:

**S1. Raman Spectroscopy**

**S2. Charge transport properties of graphene**

**S3. Control experiments in the graphene/$Bi_2O_3$ region**

**S4. Room temperature measurements**

**S5. Reproducibility in device 1**

**S6. Reproducibility in device 2**

**S7. Reproducibility in device 3**

**S8. Device 4 with Co and Au injectors**

**S9. Ordinary Hall effect measurements in graphene/Co**

**S10. Extraction of the spin transport parameters from the Hanle precession curves**

**S11. Simulations**

## S1. Raman Spectroscopy

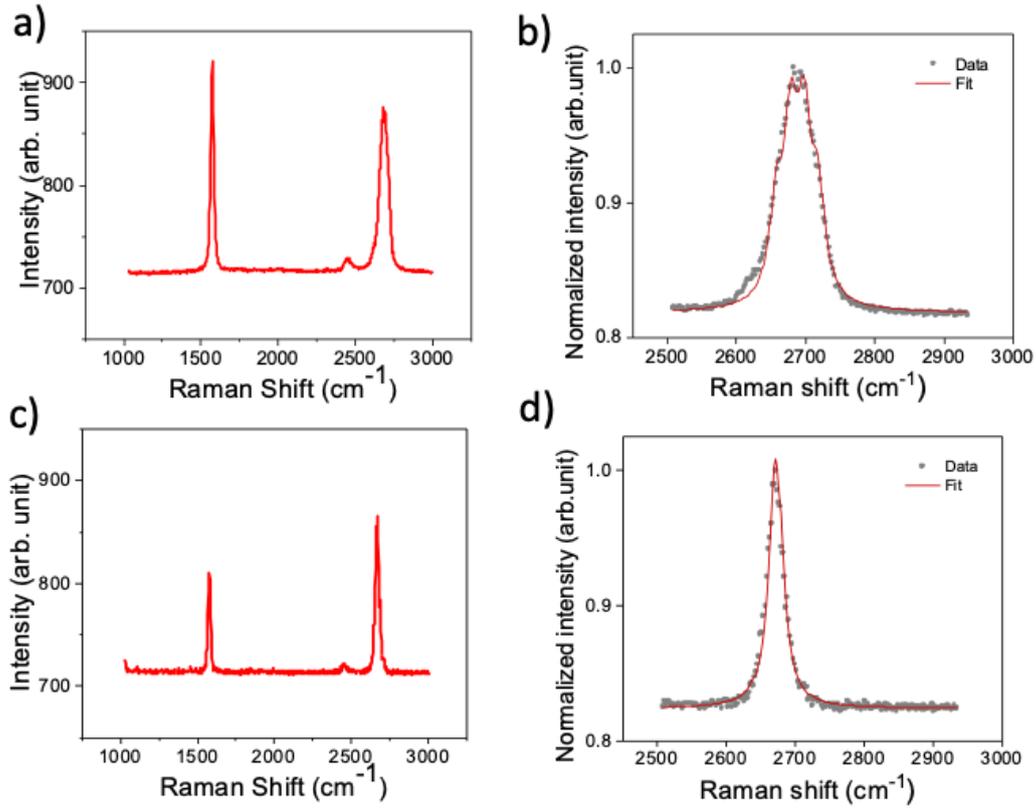

**Figure S1: a)** Raman spectroscopy of the graphene flake used in device 1 in the main text, prior to the etching process to determine the thickness of the flake. **b)** Fitting of the 2D peak into four Lorentzian functions, each with a full width at half maximum (FWHM) of ~24 cm$^{-1}$, determines the flake to be bilayer graphene [1]. **c)** Raman spectroscopy of the graphene flake used in device 3, explained in S6, prior to the etching process to determine the thickness of the flake. **d)** Fitting of the 2D peak confirming it is monolayer graphene [1].

## S2. Charge transport properties of graphene

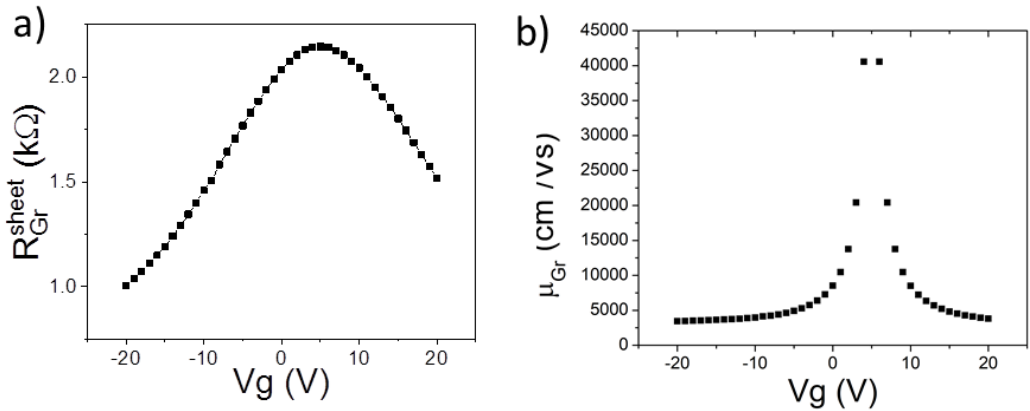

**Figure S2: a)** Sheet resistance of the pristine graphene region ($R_{Gr}^{sheet}$) as a function of $V_g$ measured at 50 K using 4-probe electrical configuration V$_{34}$I$_{AE}$ shown in figure 2a of the main text. **b)** Mobility of pristine graphene ($\mu_{Gr}$) as a function of $V_g$ calculated using the equation $\mu_{Gr} = \frac{1}{neR_{Gr}^{sheet}}$. The corresponding $R_{Gr}^{sheet}$ for each gate voltage is taken from the data in panel a.

## S3. Control experiments in the graphene/Bi₂O₃ region

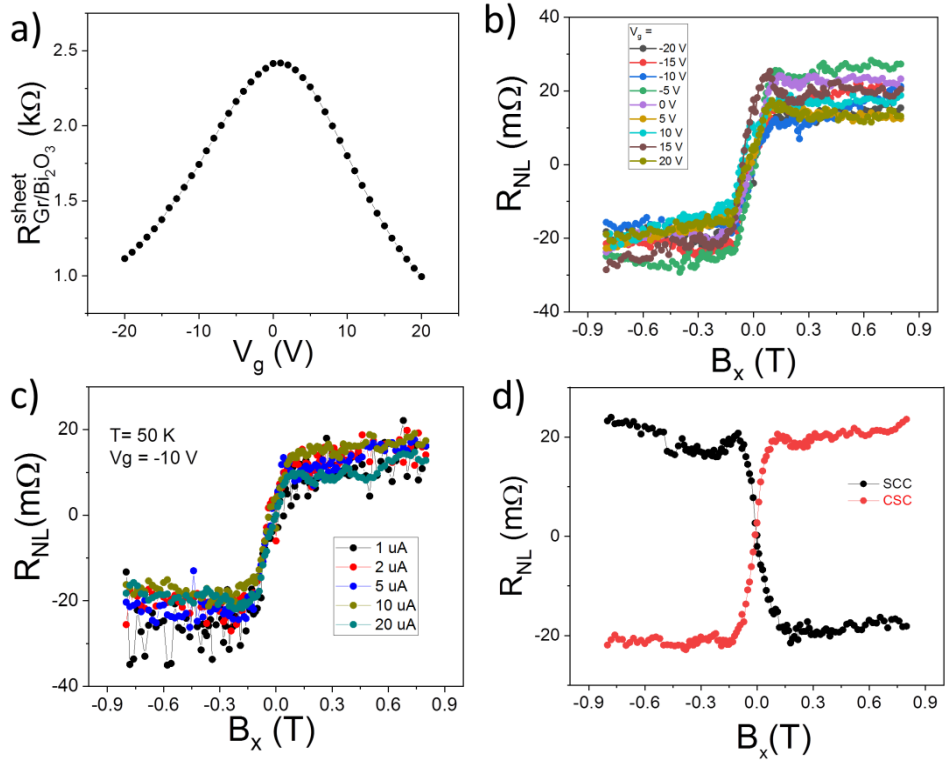

**Figure S3. a)** Sheet resistance of the graphene/Bi₂O₃ region ($R^{sheet}_{Gr/Bi_2O_3}$) as a function of $V_g$ measured using 4-probe electrical configuration V₂₃I_AE explained in figure 2a of the main text. $R_{NL}$ vs. $B_x$ measurements across the graphene/Bi₂O₃ region. **b)** for different $V_g$ from -20 V to 20 V at $I_C = 10\ \mu A$, **c)** for $I_C = 1, 2, 5, 10, 20\ \mu A$ at $V_g = -10$ V, using electrical configuration V_CDI₂A. **d)** in the SCC and charge-to-spin conversion geometries using electrical configurations V_CDI₂A and V₂AI_CD, respectively. All measurements are taken at 50 K in device 1.

## S4. Room temperature measurements

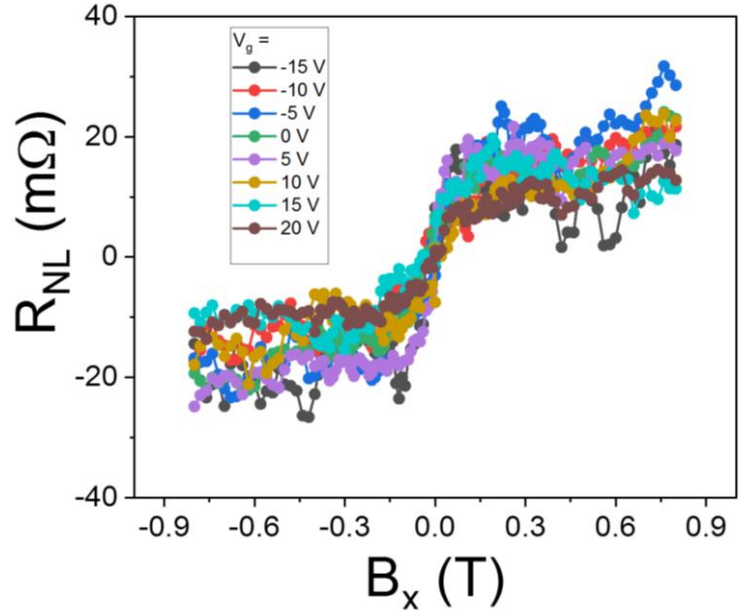

**Figure S4.** $R_{NL}$ vs. $B_x$ at room temperature obtained using the electrical configuration V_CDI₂A at different $V_g$ in device 1.

## S5. Reproducibility in device 1

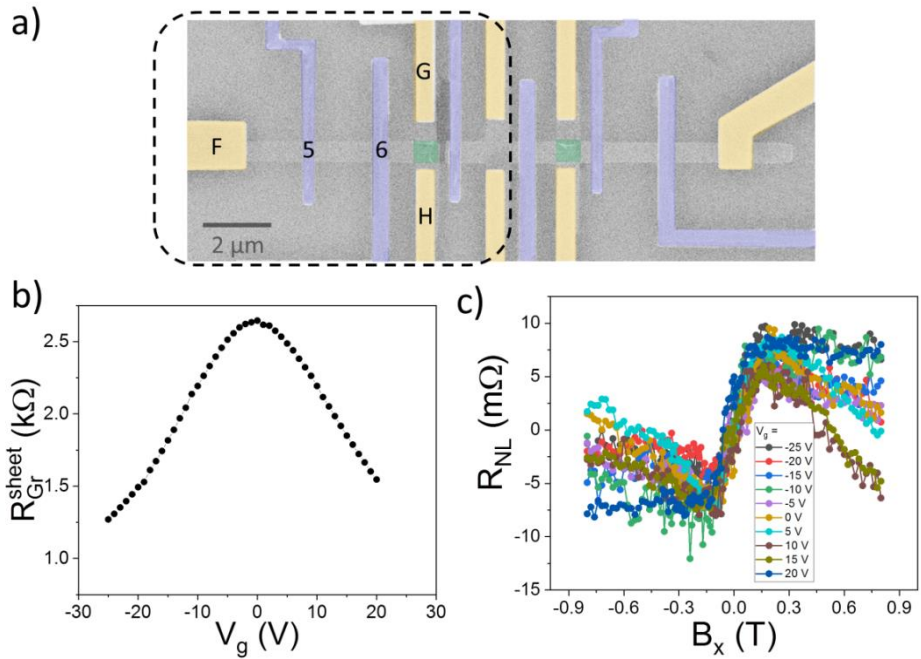

**Figure S5. a)** SEM image of the device 1 with the part of the device used for the measurements in this figure indicated by the dashed box. **b)** Sheet resistance of the pristine graphene region ($R_{Gr}^{sheet}$) as a function of $V_g$ measured using 4-probe electrical configuration $V_{56}I_{FG}$. **c)** $R_{NL}$ vs. $B_x$ measurements at 50 K across the graphene/Bi$_2$O$_3$ region for different $V_g$ from −20 V to 25 V at $I_C$ = 10 μA using electrical configuration $V_{GH}I_{6F}$.

## S6. Reproducibility in device 2

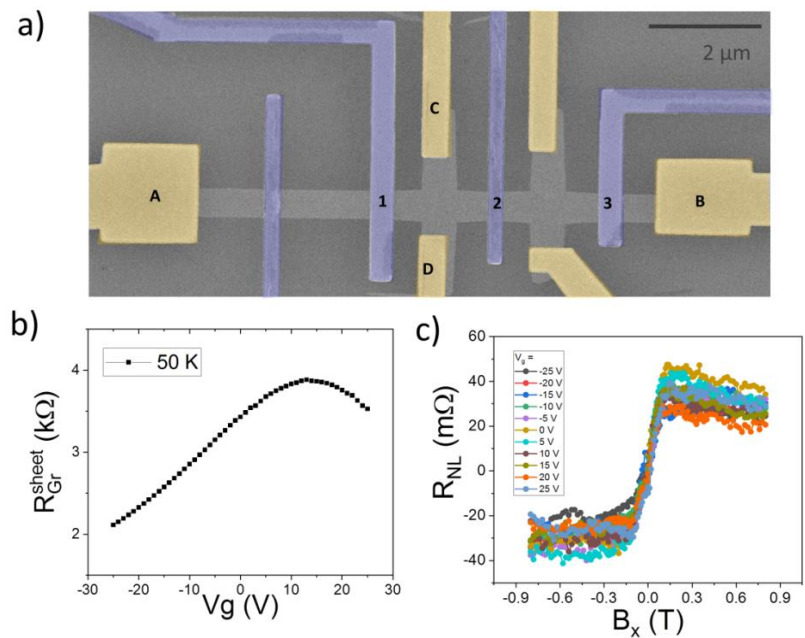

**Figure S6**. **a)** False coloured scanning electron microscope image of device 2. The Cr/Au electrodes and TiO$_x$/Co electrodes are highlighted in yellow and purple, respectively. **b)** Sheet resistance of the pristine graphene region ($R_{Gr}^{sheet}$) as a function of $V_g$ measured using 4-probe electrical configuration $V_{12}I_{AB}$. **c)** $R_{NL}$ vs. $B_x$ measurements at 50 K at the pristine graphene region for different $V_g$ from −25 V to 25 V at $I_C$ = 10 μA using electrical configuration $V_{CD}I_{1A}$.

## S7. Reproducibility in device 3

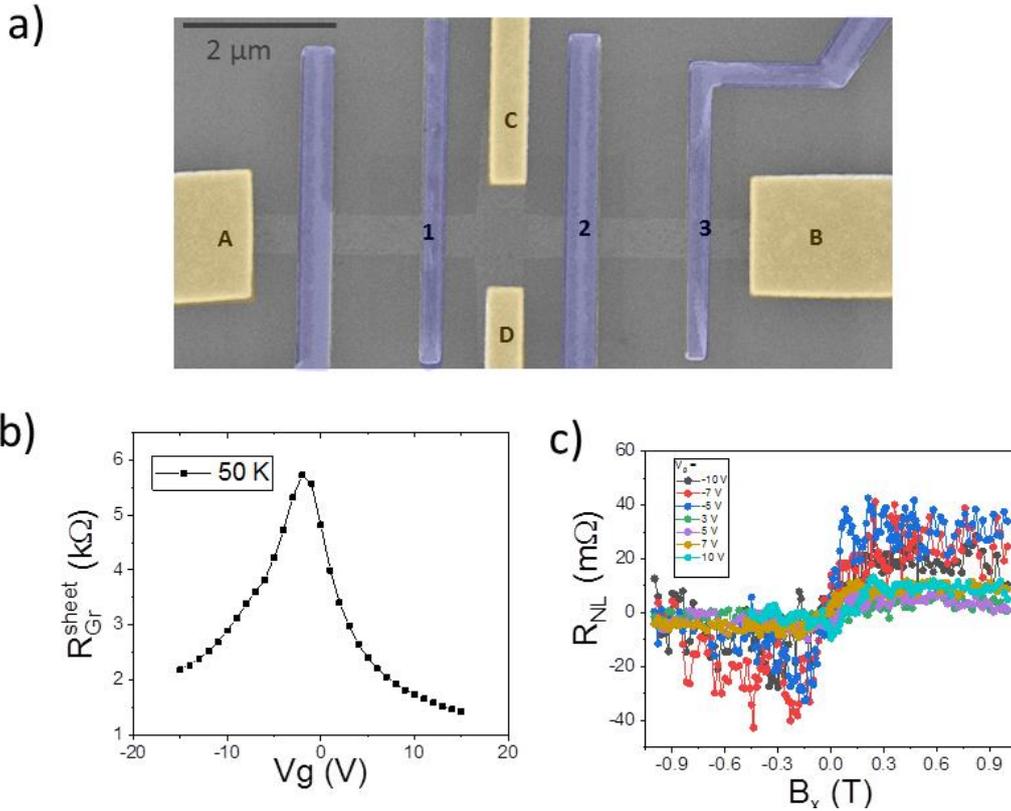

**Figure S7. a)** False coloured scanning electron microscope image of device 3 with monolayer graphene. The Cr/Au electrodes and TiO$_x$/Co electrodes are highlighted in yellow and purple, respectively. **b)** Sheet resistance of the pristine graphene region ($R_{Gr}^{sheet}$) as a function of $V_g$ measured using 4-probe electrical configuration V$_{12}$I$_{AB}$. **c)** $R_{NL}$ $vs.$ $B_x$ measurements at 50 K for different $V_g$ from −10 V to 10 V at $I_C$ = 10 µA using electrical configuration V$_{CD}$I$_{1A}$.

## S8. Device 4 with Co and Au injectors

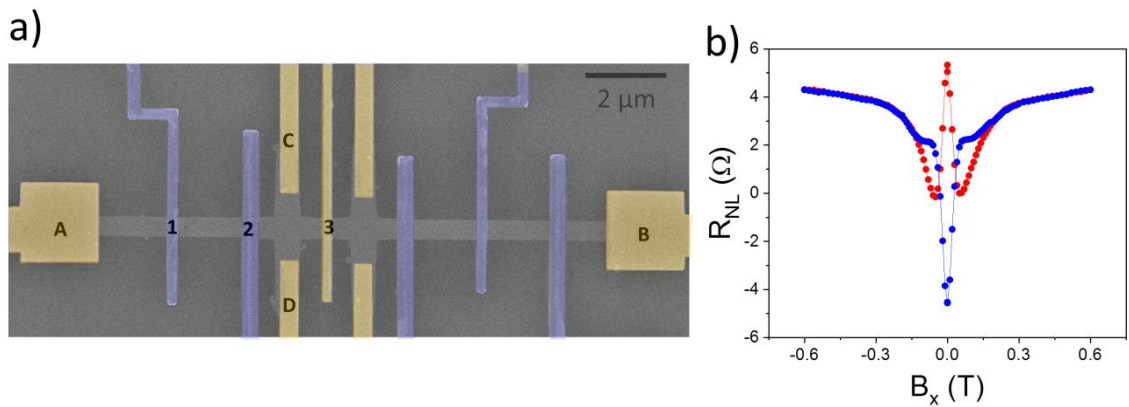

**Figure S8. a)** False coloured scanning electron microscope image of device 4. The Cr/Au electrodes and TiO$_x$/Co electrodes are highlighted in yellow and purple, respectively. The measurement in the SCC configuration is explained in figure 4a of the main text. **b)** Symmetric Hanle curves measured for initial parallel (blue curve) and antiparallel (red curve) states of the two ferromagnets across the pristine graphene region, using electrical configuration V$_{1A}$I$_{2B}$. This confirms efficient spin injection and transport in the device.

## S9. Ordinary Hall effect measurements in graphene/Co

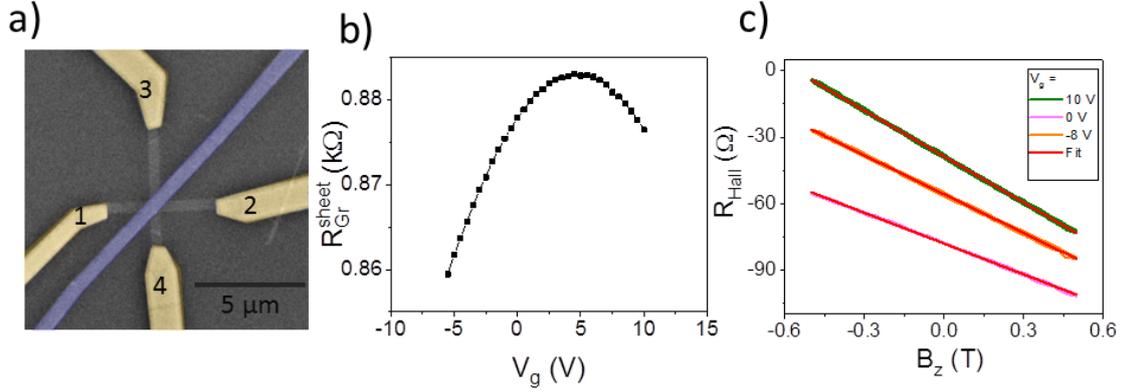

**Figure S9. a)** False coloured scanning electron microscope image of graphene Hall bar device with a Co electrode deposited on top. The Cr/Au electrodes and Co electrode are highlighted in yellow and purple, respectively. **b)** Sheet resistance of the graphene region ($R_{Gr}^{sheet}$) as a function of $V_g$ measured using 2-probe electrical configuration $V_{12}I_{12}$. Here, even though the middle part of graphene is in parallel conductance with the Co electrode on top, the contribution from the pristine graphene region to the resistance measurement is expected to be dominant. The charge neutrality point is at $V_g=4$ V. **c)** The ordinary Hall effect measurement at $V_g=10$ V, 0 V, and 8 V, respectively. Fitting the curve to the ordinary Hall effect equation, corresponding carrier densities=$9 \times 10^{12}$ cm$^{-2}$, $1 \times 10^{13}$ cm$^{-2}$, $1 \times 10^{13}$ cm$^{-2}$ were extracted. Comparing to the data in panel b, we calculate that $V_g > 140$ V will be required to achieve this carrier density in pristine graphene.

## S10. Extraction of the spin transport parameters from the Hanle precession curves

We performed nonlocal Hanle spin precession measurements using the two lateral spin valves consist of pristine graphene and graphene/Bi$_2$O$_3$ channels respectively of device 1 (between Co 2,3 and 4 in the figure 1 of the main text). To determine the spin transport parameters, we fit the measured Hanle curves using the one-dimensional spin diffusion equation governing the spin transport in the channel [2]:

$$\Delta R_{NL} = \frac{R_{NL}^P - R_{NL}^{AP}}{2} = \frac{P^2 \cos^2(\beta) R_{sq} \lambda_s}{2w} Re \left\{ \frac{e^{-\frac{L}{\lambda_s}\sqrt{1-i[g\mu_B/\hbar(B-B_0)]\tau_s}}}{\sqrt{1-i[g\mu_B/\hbar(B-B_0)]\tau_s}} \right\} \quad (1)$$

Equation 1 gives the change in non-local resistance $\Delta R_{NL}$ given by half of the difference between the measured parallel and antiparallel Hanle curves $R_{NL}^P$ and $R_{NL}^{AP}$ as a function of the magnetic field $B$, where $P$ is the spin polarization of the Co/graphene interface, $\beta$ is the angle between effective magnetization and easy axis of the ferromagnetic electrode, $R_{sq}$ is the square resistance, $w$ and $L$ are width and length of the spin transport channel and $B_0$ is a constant to compensate any small experimental offset in the measurements. The spin transport parameters spin diffusion length $\lambda_s$, time $\tau_s$, and constant $D_s$ are connected by $\lambda_s = \sqrt{\tau_s D_s}$.

We assume that the spin diffusivity $D_s$ in our sample is close to the charge diffusivity $D_c$ and calculate the latter with [3]:

$$D_c = \frac{\pi \hbar^2 v_f^2}{R_{sq} e^2 \sqrt{\gamma_1^2 + 4\pi \hbar^2 v_f^2 |n|}} \qquad (2)$$

where $v_f = 10^6$ m/s is the Fermi velocity of graphene, $\gamma_1 \sim 0.4$ eV is the interlayer coupling parameter between pairs of orbitals on the dimmer sites in BLG, $e$ the electron charge, and $\hbar$ is the reduced Planck constant.

The carrier density as a function of back gate voltage is calculated by:

$$n = \frac{\epsilon_0 \epsilon_r}{e t_{SiO_2}} (V_g - V_{CNP}) \qquad (3)$$

where $\epsilon_0$ is the vacuum dielectric permittivity, $\epsilon_r = 3.9$ is the dielectric constant of $SiO_2$, $t_{SiO_2} = 300$ nm is the thickness of the $SiO_2$ dielectric, and $V_{CNP}$ the value of the backgate voltage $V_g$ at which the graphene reaches the charge neutrality point. We determine a value for $D_s$ of $10.1 \times 10^{-3}$ m²/s that we assume is equal for both graphene and graphene/$Bi_2O_3$ regions.

The results of the fits can be seen in figure S10a and the fit parameters are given in table 1 to compare the spin diffusion parameters between the two lateral spin valves.

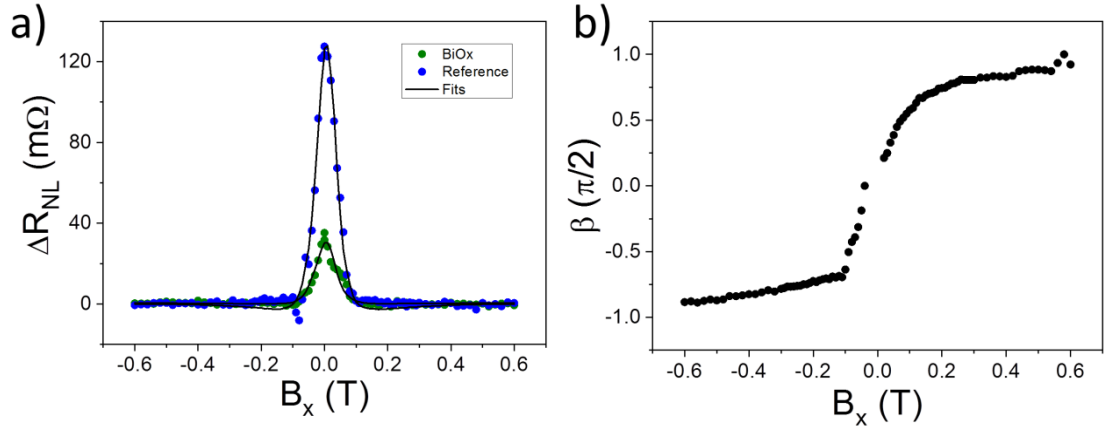

**Figure S10: a)** Measured Hanle curves as scatter plots for the lateral spin valve of pristine graphene (blue) and with $Bi_2O_3$ deposited (green) in device 1 at 50 K at $V_g = 10$ V. The black lines are a fit to Equation 1 with the resulting fit parameters given in table 1. **b)** Pulling of the magnetization of the ferromagnetic electrodes for the lateral spin valve with $Bi_2O_3$ deposited, complimentary to figure 2c of the main text.

|  | $P$ (%) | $\tau_s$ (ps) | $\lambda_s$ (µm) |
|---|---|---|---|
| Pristine graphene | 4.4 ± 0.1 | 92 ± 10 | 0.96 ± 0.10 |
| Graphene/$Bi_2O_3$ | 4.3 ± 0.2 | 60 ± 35 | 0.78 ± 0.45 |

**Table S1:** Spin transport parameters extracted from the fit of the Hanle precession curves in Figure S10a.

It should be noted that the fit of the data in the graphene/$Bi_2O_3$ region is dominated by the ferromagnetic contact pulling shown in figure S10b. The magnetization of the Co electrodes is almost completely saturated along the hard axis above 100 mT and, therefore, only a narrow field range can be used to determine $\tau_s$. This introduces a large uncertainty in the resulting fit parameter and a clear conclusion about the change of spin lifetimes between the two lateral spin valves is not possible. If there is a reduction in spin lifetime of graphene/$Bi_2O_3$ region ($\tau_s^{gr/Bi_2O_3}$) compared to that of pristine graphene ($\tau_s^{gr}$), it may indicate that $Bi_2O_3$ enhances spin-orbit coupling in graphene via proximity effect as seen in our previous report [4]. However, as the proximity effect is very sensitive to the interface properties and it strongly depends on the growth of the SOC material which can vary from device to device, the presence of $Bi_2O_3$ does not guarantee the enhancement in the SOC. With the uncertainty in the extraction of the spin lifetimes explained above, it is difficult to conclude the influence of $Bi_2O_3$ on spin transport in this study.

Finally, it should be noted that, in equation 1, the contact resistance between Co and graphene enhanced by the $TiO_2$ interface is assumed to be large compared to the sheet resistance of the graphene channel. In our case, the contact resistances of the electrodes are between 0.5 and 5 kΩ while the sheet resistance of the pristine graphene channel is 1.4 kΩ. Therefore, we underestimate the polarization of the spin injection as well as the spin lifetime of the graphene because spins will flow back into the ferromagnet due to the lower spin resistance[5,6]. This can be the reason behind the change in the amplitudes of the nonlocal spin signals between two lateral spin valves (Figure S10a).

## S11. Simulations

To quantitatively understand the influence of both the stray-field-induced OHE and the AHE in our measurements, we performed 3D simulations using finite element method (FEM) software COSMOL Multiphysics. The same device geometry consisting of a Hall bar and a Co electrode (figure S11a) is used for both calculations. The electric potential in the whole device was visualized via colour grading. Due to the complex calculation, the minimum thickness that could be used for the OHE model, which includes magnetic and electric solvers together, was limited to 8 nm. To take into account the AHE and the OHE, we used off-diagonal components of the conductance matrix which are $\pm\theta_{AHE}/\rho_{Co}$ and $\pm R_{OHE} \cdot B_z/\rho_{Gr}^2$, respectively where $\rho_{Co}$, $\theta_{AHE}$, $R_{OHE}$, $B_z$, and $\rho_{gr}$ are anomalous Hall angle, OHE resistance, out-of-plane magnetic field and graphene resistivity respectively. When a current is applied from Co to one side of the graphene channel, a corresponding local voltage is created along the current path as shown by the out-of-range blue colour in figure S11a. At the same time, a spurious current spreading towards the other side of graphene creates an electric potential giving a non-zero nonlocal voltage detected across the Hall bar. This voltage is independent of the Co magnetization causing a constant baseline voltage in the magnetic field-dependent measurements. Such baseline voltage due to van der Pauw current[6] spreading from the injecting electrode is generally observed in the graphene LSV[7] experiments, for example as shown in figure 3 of the main text.

The OHE and the AHE affect the electric potential profile creating voltages added or subtracted to the baseline voltage measured across the Hall bar. As explained in the main text, when a current is applied along the $z$ direction from Co to graphene, the AHE voltage is created in the Co electrode along the $y$ direction for the magnetization saturated along the $x$ direction. When the saturation magnetization reverses, the AHE voltage also reverses. To shunt the AHE voltage from Co to graphene, either a transparent graphene/Co interface or the presence of pinholes in the interface need to be considered. Such pinholes can be present in our real device due to the the irregular growth of the $TiO_x$ (thickness ~3 Å) barrier on top of graphene. For the case of the OHE simulations, we first calculated the out-of-plane stray field from the Co magnetization, which is shown in figure S11b. The stray field is maximum near the edge of the Co electrode and negligible at the Hall bar region. When the Co magnetization reverses, the stray field direction also reverses causing the reversal of the OHE voltage. The differences between the voltages at the Hall bar region for the Co magnetization saturated along $+x$ and $-x$ directions for both cases give the amplitude of the nonlocal signal plotted in figures 5c and 5d of the main text.

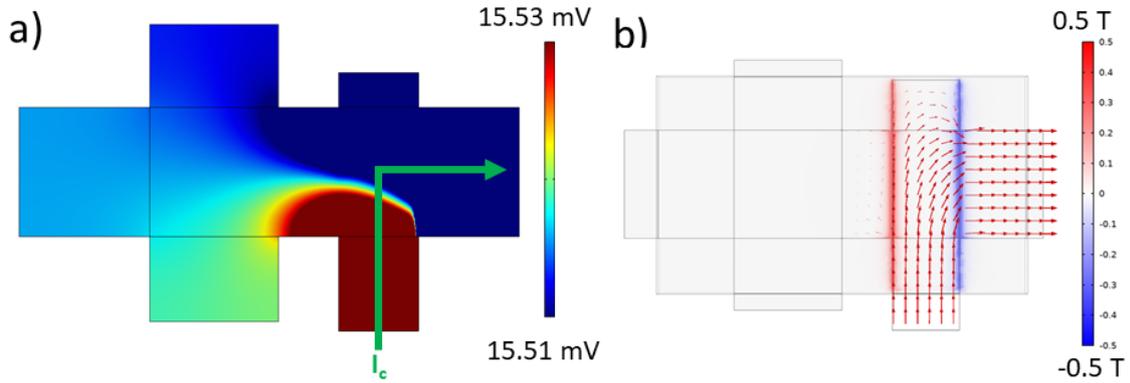

**Figure S11:** The simulations result of **a)** the electric potential profiles in the device by applying the current **b)** the stray-fields represented by colour grades in the unit of Tesla. The current path is shown by the red arrows.

For the simulation result shown in figure 5 of the main text, we used values of the graphene/Co interface resistance ($R_i$), $\theta_{AHE}$, and the number of pinholes that are suitable to the experiment condition. Those parameters might be indistinguishably different in the real device and might have affected the amplitude of the experimentally observed output signal. To understand their influence, we performed additional simulations by changing their values and the nonlocal signal ($\Delta R_{NL}$) at a detector distance of 600 nm was calculated as shown in table S2. For all the simulations, resistivities of Co ($\rho_{Co}$= 50 µΩ · cm) and graphene (($\rho_{gr}$= 162 µΩ · cm) have been used. Since the increase of $R_i$ makes the current injection into graphene more uniform, the nonlocal voltage arising from the non-uniformity was suppressed creating small variations in the output voltage. A similar dependence was also observed for the AHE output signal by varying the number of pinholes. The AHE changed linearly by varying $\theta_{AHE}$. The signal reduced exponentially by increasing distance between the edge of the ferromagnet to detector as shown in figure 5d in the main text.

Overall, as we concluded in the main text, the simulation showed the stray-field-induced OHE is the dominant mechanism contributing to the output signal as compared to the AHE in the ferromagnetic electrode.

| Parameters | Pinhole used? | $\Delta R_{NL}$ at 600 nm [m$\Omega$] |
|---|---|---|
| AHE | | |
| $R_i = 0\ \Omega$, $\theta_{AHE}$= 1% | No | 1.02 |
| $R_i = 500\ \Omega$, $\theta_{AHE}$= 1% | No | 0.37 |
| Insulating interface, $\theta_{AHE}$= 1% | Array (8*12) | 0.91 |
| Insulating interface, $\theta_{AHE}$= 1% | 2 pinholes | 0.277 |
| Insulating interface, $\theta_{AHE}$= 2% | 2 pinholes | 0.69 |
| OHE | | |
| 8nm Gr, $R_i = 1500\ \Omega$ | No | 47.754 |
| 8nm Gr, $R_i = 0\ \Omega$ | No | 58.689 |
| 16nm Gr, $R_i = 0\ \Omega$ | No | 50.752 |
| 0.81nm Gr, $R_i = 0\ \Omega$ | No | 65.8 |

**Table S2:** The net output signal ($\Delta R_{NL}$) measured at detector distance = 600 nm shown in column 3 by varying different parameters as shown in column 1 and 2 ($R_i$, $\theta_{AHE}$, thickness of graphene and the number pinholes)